**Auditory cortex and beyond: Deficits in congenital amusia**

Barbara Tillmann (1,2,*), Jackson E. Graves (3), Francesca Talamini (4), Yohana Leveque (1), Lesly Fornoni (1), Caliani Hoarau (1), Agathe Pralus (1), Jérémie Ginzburg (1), Philippe Albouy (5) & Anne Caclin (1,*)

[1] Université Claude Bernard Lyon 1, CNRS, INSERM, Centre de Recherche en Neurosciences de Lyon CRNL, UMR5292, U1028, F-69500, Bron, France

[2] Laboratory for Research on Learning and Development, LEAD – CNRS UMR5022, Université de Bourgogne, Dijon, France

[3] Laboratoire des systèmes perceptifs, Département d'études cognitives, École normale supérieure, PSL University, CNRS, 75005 Paris, France

[4] Department of Psychology, University of Innsbruck, Austria

[5] CERVO Brain Research Center, School of Psychology, Laval University, Québec, G1J 2G3; International Laboratory for Brain, Music and Sound Research (BRAMS), CRBLM, Montreal, QC, H2V 2J2, Canada

* Corresponding authors: Barbara Tillmann & Anne Caclin, CNRS UMR5292-INSERM U1028, Lyon Neuroscience Research Center, 69675 Bron Cedex, France. E-mail address: Barbara.Tillmann@cnrs.fr & anne.caclin@inserm.fr

Acknowledgements: This work was conducted in the framework of the LabEx CeLyA ("Centre Lyonnais d'Acoustique", ANR-10-LABX-0060) and of the LabEx Cortex ("Construction, Function and Cognitive Function and Rehabilitation of the Cortex", ANR-11-LABX-0042) of Université de Lyon, within the program "Investissements d'avenir" (ANR-11-IDEX-0007) operated by the French National Research Agency (ANR).

**HIGHLIGHTS**

Congenital amusia is linked to altered pitch processing and short-term memory = 77
Impairments extend to the processing of pitch cues in speech and emotion = 72
Linked to altered brain responses in a distributed fronto-temporal network = 74
Implicit processing  and audio-visual integration open perspectives of rehabilitation = 85

**Abstract**

Congenital amusia is a neuro-developmental disorder of music perception and production, with the observed deficits contrasting with the sophisticated music processing reported for the general population. Musical deficits within amusia have been hypothesized to arise from altered pitch processing, with impairments in pitch discrimination and, notably, short-term memory. We here review research investigating its behavioral and neural correlates, in particular the impairments at encoding, retention, and recollection of pitch information, as well as how these impairments extend to the processing of pitch cues in speech and emotion. The impairments have been related to altered brain responses in a distributed fronto-temporal network, which can be observed also at rest. Neuroimaging studies revealed changes in connectivity patterns within this network and beyond, shedding light on the brain dynamics underlying auditory cognition. Interestingly, some studies revealed spared implicit pitch processing in congenital amusia, showing the power of implicit cognition in the music domain. Building on these findings, together with audio-visual integration and other beneficial mechanisms, we outline perspectives for training and rehabilitation and the future directions of this research domain.

**Keywords**

1. <u>Introduction</u>

Congenital amusia is a neuro-developmental disorder of music perception and production that has increasingly attracted cognitive neuroscience research over the last two decades (see Peretz, 2016; Tillmann et al., 2015 for reviews), partly thanks to a widely accepted screening tool (i.e., the Montreal Battery of Evaluation of Amusia (MBEA), Peretz et al., 2003). Its prevalence has been initially estimated to 4% (e.g., Kalmus & Fry, 1980; Peretz et al., 2003), but more recently revised to 1.5 % (Peretz & Vuvan, 2017). Congenital amusia has been compared to other neuro-developmental disorders, such as prosopagnosia or dyslexia (e.g., Hyde et al., 2006; Peretz et al., 2002), and its potential comorbidity with dyslexia has been investigated (e.g, Couvignou et al., 2019, 2023; Couvignou & Kolinsky, 2021). It could be compared to an apperceptive agnosia in the auditory modality, but as we will see below, this condition includes a strong short-term memory component. Its origin is hypothesized to be linked to genetics, as suggested by family aggregation studies in particular (e.g., Peretz et al., 2007; Peretz & Vuvan, 2017). Importantly, the deficit cannot be explained by peripheral hearing loss or brain lesions, by social or more general cognitive deficits (e.g., Ayotte et al., 2002; Peretz et al., 2002), with for example normal performance in classical digit span tasks (e.g., Albouy, Schulze, et al., 2013; Tillmann et al., 2009; Williamson, McDonald, et al., 2010). As anecdotal support, examples of highly educated amusic individuals have been reported (e.g., Che Guevara, Milton Friedman, see Peretz, 2003; Münte, 2002).

While the findings of an increasing number of research studies reveal the possibility of various forms of congenital amusia, similar to the various forms that exist for acquired amusia (with accidental brain damage), the main dimension affected by this disorder seems to be the pitch dimension, which has also been the focus of most of the investigations. Variabilities seem to exist as to whether processing of the time dimension is also impaired (e.g., Foxton et al., 2006; Pfeuty & Peretz, 2010), as well as regarding potential enjoyment of music listening (e.g., Mcdonald & Stewart, 2008; Omigie et al., 2012). Furthermore, most amusic participants, but not all, exhibit poor singing (Dalla Bella et al., 2009). Pitch perception and production can be dissociated to some extent in this disorder (Loui et al., 2008).

The main hypotheses for impairments underlying the condition of congenital amusia are impaired fine-grained pitch perception (e.g., Ayotte et al., 2002; Foxton, Dean, et al., 2004; Hyde & Peretz, 2004) and impaired short-term memory for pitch, which is observed even in the absence of elevated pitch discrimination thresholds or when the pitch changes to be processed exceed amusics' individual pitch discrimination thresholds (e.g., Gosselin et al., 2009; Tillmann et al., 2009, 2016; Williamson, Baddeley, et al., 2010; Williamson & Stewart, 2010).

Congenital amusia is typically diagnosed when a participant's score falls below some threshold at the MBEA (Peretz et al., 2003). It is noteworthy that the MBEA does not test amusia as a "construct" but rather evaluates multiple (but not all) dimensions related to music perception and cognition. The MBEA contains six sub-tests addressing notably the pitch (or melodic) dimension (corresponding to the detection of an out-of-key note, a contour violation, or interval change), the time dimension (rhythm and meter perception), and incidental memory (i.e., memory of melodies used in preceding sub-tests). Complementary subtests and tools aiming for a more robust screening of amusia in the general population have been proposed. These tools notably test harmony and emotion (Sloboda et al., 2005, subtests that can be added to the MBEA: MBEA(R)), pitch discrimination threshold (Tillmann et al., 2009), time discrimination threshold (van Vugt et al., in preparation), self-evaluation/reports and questionnaires (Wise and Sloboda, 2008; Tillmann et al., 2023; Mcdonald & Stewart, 2008; Omigie et al., 2012, Peretz et al., 2008), and vocal production (Wise and Sloboda, 2008). Vuvan, Paquette et al. (2018) recently proposed a more complete screening protocol for both music perception and production, completed with a questionnaire. Other research groups have used different tests to define a population with a musical deficit. Notably, the Distorted Tune Test (DTT) investigates participants' capacity to discriminate intervals between tones in familiar melodies (Drayna et al., 2001; Jones, Lucker et al., 2009; Jones, Zalewski et al., 2009; Kalmus and Fry, 1980). This test uses well-known tunes (in North America) and is dependent on participants' long-term memory knowledge of a cultural musical repertoire. As the MBEA is i) based on newly composed melodies, ii) evaluates several (but not all) aspects of music perception, iii) contains catch trials that are not considered in the score (to rule out individuals who are responding randomly or do not understand the task instructions), iv) has been used successfully across various countries and cultures (North America, Europe (e.g., France, UK,

Greece), China, and New Zealand), this battery has thus been well adopted by the research community.

In the present article, we focus on the pitch dimension and review research aiming to understand this phenomenon with its deficits, anomalies, but also its spared functioning, leading us to outline perspectives for training or rehabilitation. Studying congenital amusia provides a unique opportunity to increase our understanding of typical and pathological human auditory network functioning, in particular here pitch perception and memory (Section 2). The pitch processing impairment applies in particular to musical material (i.e., for which it is a form-bearing dimension, e.g., McAdams, 1989), but affects also processing of musical timbres and speech material, if pitch is relevant, such as for intentional or emotional prosody or in tone languages (Section 3). Structural and functional brain imaging data of the amusic brain have shown impairments in fronto-temporal networks as well as the connectivity within and between the auditory cortices during pitch encoding and memory (Section 4). A set of research using implicit, indirect testing methods have revealed some preserved implicit processing mechanisms and musical knowledge (e.g., the tonal structures of the musical system, specific musical pieces), leading to the hypothesis of a potential disorder of consciousness (Section 5). Building on these findings of implicit perception and cognition together with others showing the potential of cross-modality boosting of pitch processing, new research perspectives start emerging, aiming to design directions for training and rehabilitation of this disorder (section 6). Keypoints on congenital amusia as reviewed here are summarized in Inserts 1 to 3.

## 2. From the hypothesis of impaired pitch discrimination to that of impaired memory

The seminal papers describing congenital amusia (Ayotte et al., 2002; Peretz et al., 2002) emphasized the music specificity of the disorder, with intact speech and environmental sound processing. They also pinpointed impaired pitch processing as a possible candidate to explain the musical deficits. In Peretz et al. (2002), the extensive testing of a single case of congenital amusia revealed deficits in pitch discrimination, pitch change detection, melody comparison (short-term memory), detection of out-of-key notes in melodies (tonality processing), and familiar melody recognition (long-term memory). The first group study of congenital amusia revealed deficits in melody comparison (short-term memory), deficits in detecting out-of-key notes in melodies (tonality), reduced sensitivity to dissonance, difficulties in explicitly recognizing the melody of famous songs (long-term memory), and poor singing (Ayotte et al., 2002). Based on these first studies (see also Foxton, Brown, et al., 2004), it was hypothesized that an impoverished fine-grained pitch perception was the core deficit in congenital amusia. This hypothesis was tested by Hyde and Peretz (Hyde & Peretz, 2004) using a pitch change detection task and a matched time change detection task. Participants were presented with sequences of five isochronous tones of constant pitch, where the fourth tone could be changed in pitch or displaced in time. Congenital amusics (identified with the MBEA) performed worse than controls for the pitch task, but not for the time task, thus lending support to the hypothesis of a primary impairment in pitch processing, sparing temporal processing. The deficit was mostly observed for small pitch changes (i.e., 25 and 50 cents), in keeping with the fine-grained pitch processing deficit hypothesis. While these original studies have used fixed pitch changes, numerous studies have since confirmed the pitch processing deficit in congenital amusia, with various tasks and adaptative procedures (e.g., Albouy, Schulze, et al., 2013; Foxton, Dean, et al., 2004; Liu et al., 2010; Stewart, 2011, Tillmann et al., 2009, van Vugt et al., in preparation). It is important to note that this pitch deficit was not restricted to musical contexts, but was also observed with relatively simple acoustic material (e.g., Foxton, Brown, et al., 2004; Hyde & Peretz, 2004; Tillmann et al., 2015).

Deficits that are less pronounced than those on the pitch dimension have been reported for the rhythm and/or meter subtests of the MBEA (starting with the studies of Ayotte et al., 2002; Peretz et al., 2002; see also Vuvan et al., 2018; Figure 1 in the present paper). Recently, we adapted the task of Hyde and Peretz (2004), to measure pitch and time discrimination thresholds (with an adaptative procedure) and tested a larger sample size. Our findings confirmed not only a deficit on the pitch dimension in congenital amusia, but revealed also a deficit on the time dimension, although less pronounced than the pitch deficit (van Vugt et al., in preparation). It has been proposed that time processing deficits are more likely to arise in congenital amusia when the sound material entails pitch variations. In a short-term memory task requiring the comparison of two sequences that might be altered in their rhythmic pattern, performance of the amusic group was below that of the control group when the sequences were composed of tones varying randomly in pitch, but not for monotonic sequences (note however that there was a statistical trend towards group differences also in the condition without pitch variation (Foxton et al., 2006)). In a duration discrimination task, Pfeuty & Peretz (2010) did not observe a difference between amusics' and controls' performance in a pitch-constant condition. Yet other studies suggested spared beat processing in congenital amusia (Phillips-Silver et al., 2013), but also with an influence of the pitch dimension: rhythmic performance of amusic participants was not different to that of controls when using non-pitched (drums) whereas it was impaired with pitched (e.g., tones) stimuli. Overall, time processing and rhythm processing might be impaired, but to a lesser extent than pitch processing in congenital amusia, with further performance impairments when the material entails pitch variation. In parallel to research on the pitch-based form of congenital amusia, another line of evidence has revealed that time and rhythm processing can be selectively impaired (e.g., Launay, et al,, 2014), including beat deafness (e.g., Mathias et al., 2016; Tranchant & Peretz, 2020). Future work should aim at clarifying when and how pitch and time deficits can be associated in congenital amusia and other neurodevelopmental disorders (see also Lagrois & Peretz, 2019; Couvignou et al., 2023).

Several studies have explored the nature of the pitch deficit in amusia. Pitch is generally the perceptual correlate of fundamental frequency (F0), but it is known to interact with brightness, an aspect of timbre which is the perceptual correlate of the centroid of the spectral envelope (e.g., Allen & Oxenham, 2014; Melara & Marks, 1990), with pitch and brightness perhaps relying on overlapping neural mechanisms (e.g., Allen et al., 2017, 2019). In amusia, it has also been suggested that pitch deficits may affect this "spectral pitch" mechanism, leading to impairments in the short-term memory of timbre sequences (Marin et al., 2012; Tillmann et al., 2009). Timbre perception might thus be also affected through brightness. Indeed, deficits in short-term processing of pitch contour appear to extend to brightness contours, but not loudness contours (Graves et al., 2019), suggesting that this shared spectral mechanism may be affected in amusia. Further support for the involvement of the spectral code comes from the finding that amusics are unimpaired for pitch perception of harmonic complexes with only unresolved components (Cousineau et al., 2015), a task that likely involves extracting F0 using temporal-envelope cues, since spectrum-based cues are unavailable. However, there is also some evidence that detection of amplitude modulation (AM), an important mechanism for temporal envelope extraction, which is not usually thought to be involved with pitch perception, is impaired in amusia. An early study suggested that the perception of roughness, produced by AM beats between interacting frequency components, is intact in amusia (Cousineau et al., 2012), but further studies found behavioral impairment in amusics for low levels of AM detection, near threshold (Graves et al., 2023; Whiteford & Oxenham, 2017). Future research should attempt to clarify whether temporal envelope perception is fully intact in amusia, or whether elevated AM thresholds in amusia may lead to impaired temporal envelope perception in more adverse conditions.

Congenital amusia is thus understood as a pitch (and timbre) processing deficit. Beyond simple discrimination tasks and discrimination threshold measurements, it was observed, some years after the initial description of congenital amusia, that pitch short-term memory was impaired in this condition (Gosselin et al., 2009a, 2009a; Tillmann et al., 2009a; Williamson, Baddeley, et al., 2010b; Williamson, McDonald, et al., 2010). Pitch short-term memory has been mostly assessed with recognition tasks, usually delayed-matching-to-sample tasks.  Importantly, the pitch short-term memory deficit was observed even when the pitch changes involved in the memory task exceeded individual pitch discrimination thresholds (Tillmann et al., 2009a; Williamson & Stewart, 2010). In keeping with these findings, pitch short-term memory performance of amusics stands well below that of controls at the individual level in different task settings (see for example Figure 2), whereas pitch discrimination thresholds can be in the typical range, even if they are elevated at the group level, in comparison to controls (Figure 1D). As illustrated in Figures 1E and 1F, pitch short-term memory performance (approximated by the MBEA pitch score, that is the average of the three subtests of the MBEA entailing melody short-term memory processes) is largely independent of pitch discrimination abilities (measured with an adaptive threshold task, Tillmann et al., 2009). The pitch short-term memory deficit of congenital amusics is exacerbated by factors known to be deleterious to memory processing (review in Tillmann et al., 2016): increasing the retention delay (Williamson, McDonald, et al., 2010); increasing the number of items to memorize, i.e., the memory load (Gosselin et al., 2009); and presenting interfering tones during the maintenance phase (Gosselin et al., 2009; Williamson & Stewart, 2010). However, pitch short-term memory in congenital amusia can be improved by using structured tonal material, compared to atonal material, similarly to what is observed in controls as well as for other structured materials in other domains (Albouy, Schulze, et al., 2013b; Lévêque et al., 2022). Investigating short-term memory in the amusic population provides thus a complementary view of its investigation in typical functioning for various aspects, whether detrimental or beneficial.

To reconcile data on pitch discrimination and pitch short-term memory deficits in congenital amusia, it has been proposed that the core deficit concerns pitch short-term memory, and that when this memory deficit is strong enough then even performance in basic discrimination tasks with very limited retention times can be affected (see Tillmann et al., 2016). It is worth noting that memory tasks and basic psychoacoustic tasks share the encoding stage in short-term memory because the sequential presentation of items in psychoacoustic tasks also requires encoding of pitch in memory, even though followed by a shorter maintenance duration than that in short-term memory tasks.

Pitch encoding has been shown to take a few hundred milliseconds, and thus to benefit from slow paces of sound presentation allowing to build the pitch representation (Demany & Semal, 2005). Interestingly, it was found that congenital amusic participants are especially impaired relative to controls with fast presentation rates (i.e., SOA of 100ms), both in pitch sequence short-term memory tasks and two-tone discrimination tasks (Albouy et al., 2016). This result suggests a memory deficit in congenital amusia arising as early as during the pitch encoding stage, with consequences in a variety of tasks. This result suggests a memory deficit in congenital amusia arising as early as during the pitch encoding stage, with consequences in a variety of tasks. Indeed, auditory memory requires several processing steps, notably after the extraction of the auditory attributes by the perceptual system, the information needs to be maintained in echoic memory where a pitch memory trace of the sound is established and then stored in auditory short-term memory for several seconds or minutes. In amusia, the deficit might start with impairments in the first processing steps, including an impaired pitch encoding that leads to a fragile or impaired memory trace (see Albouy et al., 2016 for further discussion).

Importantly, the memory deficit in congenital amusia cannot be explained by a general memory deficit in this condition. Indeed, short-term memory for verbal information is preserved, as reported for both delayed-matching-to-sample tasks (Tillmann et al., 2009), and classic digit span tasks (Albouy, Schulze, et al., 2013; Williamson, McDonald, et al., 2010).

3. <u>Impaired pitch processing affecting emotion processing and speech processing</u>

Pitch processing is not only a key element of musical structure processing, but also of musical emotion processing (e.g., Schellenberg et al., 2000) and speech processing (Bowles et al., 2016; Marin, 2018; Nooteboom, 1997; Pihan, 2006), notably for emotional prosody, intentional prosody, and tone languages.

For music, different studies investigated whether amusics have spared or impaired identification of the emotions conveyed (e.g., judging whether a musical piece is happy or sad) and the level of intensity of these emotions (Fernandez et al., 2020; Gosselin et al., 2015; Lévêque et al., 2018). Results appear to be mixed, with some studies reporting impaired musical emotion recognition in amusic individuals (Lévêque et al., 2018; Zhou et al., 2019), while others did not observe this deficit (Fernandez et al., 2020; Loutrari & Lorch, 2017) or observed only a reduced use of some available tonal cues, such as mode (Gosselin et al., 2015). Similarly, the judgment of intensity of musical emotions has been reported to not differ from that of controls in the study of Lévêque et al. (2018), suggesting intact implicit processing of musical emotions, but to differ from controls in the study of Fernandez et al. (2020). It has to be noted that the experimental designs used in these studies are heterogeneous, ranging from direct evaluations of emotions (e.g., sad vs. happy) of minor vs. major chords (e.g., Zhou et al., 2019), to the categorization of complex melodies in more nuanced emotional categories or of full orchestral musical excerpts (e.g., joy, tension, tenderness, sadness, Fernandez et al., 2020; Lévêque et al., 2018). These data patterns suggest that amusics might be impaired in recognizing musical emotions under certain conditions, which needs to be further investigated in future research. Impairments in the processing related to tonal cues (e.g., mode), pitch or spectral information seem to be one reason leading to emotional processing deficits. These deficits might also be (at least partly) compensated by the use of other features, such as temporal features (tempo, pulse clarity) or timbre (e.g., Gosselin et al., 2015).

Pitch processing is also relevant for identifying the emotions and intentions conveyed by speech prosody (Nooteboom, 1997; Tang et al., 2017). Several studies have investigated whether amusics' pitch perception and pitch memory deficit might also affect prosody perception (e.g., Cheung et al., 2021; Lolli et al., 2015; Pralus et al., 2019; Thompson et al., 2012). In the study by Thompson et al. ( 2012), amusics and matched control participants were presented with spoken sentences that were semantically neutral, but conveyed four different emotions (happy, tender, afraid, irritated, sad) or no emotion (i.e., neutral), which had to be identified by the participants. Results showed that amusics had significantly lower performance in correctly identifying the emotions conveyed by the prosody, and the impairment was more prominent in sentences that had similar intensity and duration. This finding suggests that the pitch processing impairment also affects speech processing, except when other-than-pitch cues are present, such as cues related to intensity or duration, allowing amusics for correctly identifying the emotions conveyed by the prosody. The difficulty of amusics in correctly identifying the emotional prosody in sentences with neutral semantic content was also reported by others (Cheung et al., 2021; Lima et al., 2016). This difficulty cannot be reduced to a more general impairment in emotion recognition, as amusics were as accurate as controls in identifying emotions that were semantically conveyed by the semantics of written words (Cheung et al., 2021) or by faces (Lévêque et al., 2018).

Some studies reported that amusics performed as well as their controls when categorizing the emotion of spoken sentences (with two emotion categories: Lolli et al., 2015; Loutrari & Lorch, 2017; with more than two emotion categories: Pralus et al., 2019). In these cases, amusics might benefit not only from cues other than pitch, but also from accumulating evidence when listening to longer sentences. Indeed, amusics were impaired in emotion recognition for isolated vowels, in particular for the distinction between sad and neutral stimuli, whereas in the same study they were unimpaired with full sentences (Pralus et al., 2019). This lower performance was linked with amusics' difficulties in processing pitch and spectro-temporal parameters of the vowels. In a recent study, we used these same vowels conveying different emotions in a passive listening paradigm to study preattentive processing, assessed via EEG recordings (Pralus et al., 2020). Results showed some emotion-specific differences between the groups in EEG components for the emotional oddball in a sequence of vowels. For example, a decreased early negative component emerged for neutral and sad vowels in amusics, together with a decreased amplitude of P3a for the vowel representing anger. These results suggest that even though amusics seem to be able to implicitly detect a change in the emotion conveyed by speech prosody, they might be impaired in early processing of the related acoustical changes. Similarly as for musical emotion perception, future studies should clarify the conditions under which emotional prosody perception might be impaired or spared, including potential inter-individual differences among amusic profiles as well as their extent of pitch processing deficit.

Pitch is also a relevant cue for intentional prosody in speech. While first studies reported amusics to be unimpaired in classifying spoken sentences as statement or question (based on the final pitch information, which might be rising or falling) and in identifying or discriminating stressed words in sentences (e.g., Ayotte et al., 2002; Patel et al., 2005; Peretz et al., 2002), more recent studies also revealed deficits for intentional prosody in amusia. In comparison to controls, amusics were impaired for the processing of speech intonation (question vs. statement) in their native language (English or French; Liu et al., 2010; Patel et al., 2008), even when comparing directly two spoken words (Lu et al., 2015) and, in particular, when more fine-grained pitch differences are implemented in the material (Hutchins et al., 2010). The pitch processing deficit thus extends to speech processing (Vuvan et al., 2015). This has been observed also for controlled experimental material with sequences of syllables (and tones in comparison). Using a pitch change detection paradigm (as in Hyde & Peretz, 2004), the detection of a pitch change in a syllable sequence (i.e., the repeated presentation of the spoken syllable /ka/) was impaired in congenital amusic individuals. Interestingly, this pitch change detection impairment was less strong for the syllables than for the tones, in particular for amusics with large pitch discrimination thresholds (Tillmann, Rusconi, et al., 2011). Jasmin and collaborators (Jasmin et al., 2020) asked participants to match a speech stimulus containing intentional prosody differences (on pitch and/or duration cues) with a written sentence. For amusics, they reported altered functional connectivity between left prefrontal language-related regions and right hemisphere pitch-related regions when performing the task.

Pitch is not only a central element of music (for both structure and emotion) and of speech prosody, but also of tone languages (see Marin, 2018, for a review). In tone languages (e.g., Mandarin), the meaning of words can change with the intonation of a syllable, and this intonation is largely determined by pitch variations. It seems thus relevant to investigate whether amusics' pitch perception and pitch memory deficit might impact tone language perception. Nan et al. (2010) showed that the phenomenon of congenital amusia is also observed among tone language speakers. About half of Mandarin-speaking amusics had some impairments in the discrimination and identification of lexical tones, with some showing such strong impairments that they qualify for lexical tone agnosia (for perception, but not for production). The deficits in the processing of pitch contrasts in tone language words have been shown not only for native speakers (e.g., Mandarin speakers, Jiang et al., 2010, 2012; Nan et al., 2010), but also for non-native speakers (e.g, French speakers, Nguyen et al., 2009; Tillmann, Burnham, et al., 2011). Impairments in tone discrimination for Mandarin-speaking amusic individuals have also been observed at a pre-attentive processing level, notably with reduced MMNs in comparison to their matched controls (Nan et al., 2016). In everyday life, native speakers of tone languages who are amusics might not encounter speech perception difficulties because other acoustic attributes can co-vary with pitch information in the tones (e.g., duration cues, intensity cues) and together with semantic or contextual cues these are helpful for perception and tone identification. The association between amusia and lexical tone processing deficits in the lab suggests that the pitch impairments are not restricted to musical material, but more generally affect auditory processing, including speech.

4. Fronto-temporal pathway anomaly in congenital amusia

In contrast to acquired amusia, congenital amusia is a phenomenon occurring without brain damage or brain lesions. Numerous neuroimaging research has however revealed that this condition is related to cortical anomalies at both structural and functional levels. The findings were reported for passive listening, for listening tasks on the target dimension (i.e., pitch) or a different dimension, as well as during resting state. They provided converging evidence for functional and structural anomalies on the fronto-temporal pathway (Figure 3), which has been previously reported to be implicated in pitch perception and memory in neurotypical participants (see for example, the seminal data of Zatorre et al., 1994; see also Schulze et al., 2011).

4.1. *Structural* brain anomalies

Structural brain anomalies were first revealed in congenital amusia using Voxel-Based Morphometry (VBM, Hyde et al., 2006) described a reduction in white matter concentration in the right Inferior Frontal Gyrus, which correlated with performance in the MBEA Scale subtest. This anomaly in the right IFG was also observed with VBM in Albouy et al. (Albouy, Mattout, et al., 2013). The study of cortical thickness revealed increased grey matter in the right IFG and in the right auditory cortex (Hyde et al., 2007) in congenital amusia, both correlating with MBEA scores. Besides these subtle focal changes, a reduction in the connectivity between the right auditory and frontal cortices was observed using Diffusion Tensor Imaging (DTI), in particular a reduced right arcuate fasciculus (AF) in congenital amusia (Loui et al., 2009). Correlations were observed between pitch discrimination and the volume of the superior branch of AF, and between an index of pitch production and the volume of the inferior branch of AF. This finding of a reduced anatomical pathway between auditory temporal and frontal areas received further support in subsequent DTI studies (Chen et al., 2018; Wang et al., 2017), yet these findings were not replicated by Chen et al. (2015). In addition, DTI studies revealed decreased whole-brain connectivity scores in congenital amusia (Wang et al., 2017; Zhao et al., 2016), which might however be driven by the fronto-temporal anomaly (Wang et al., 2017).

*4.2. Functional* brain anomalies

In addition to anatomical anomalies and their link to behavioral correlates, a set of studies have investigated the functional correlates of pitch perception and memory in congenital amusia (using EEG, MEG, and fMRI, see also section 5). The first fMRI study investigating amusics' functional neural correlates used a passive listening paradigm, with pure-tone melody-like patterns differing in the size of pitch variations (Hyde et al., 2011). The findings provided converging evidence for an altered fronto-temporal network in amusia, here on a functional level. Indeed, amusic participants showed decreased activation of the right IFG as well as a decreased right fronto-temporal functional connectivity between the right IFG and the right STG (in comparison to the controls). While these first fMRI findings did not show a deficit in the auditory cortex, they revealed an over-connectivity between the two auditory cortices in amusic participants.

Using an active perception task, Norman-Haignere et al. (2016) set out to further investigate a potential anomaly of the auditory cortex in amusia. Participants listened to sequences of harmonic tones (and acoustically matched noise as control condition) and performed an active task on temporal aspects of the stimuli. Focusing the analyses on the pitch-responsive regions in the auditory cortex did not reveal any differences between amusic and control participants, neither in extent, nor selectivity nor anatomical location. This result pattern was also valid when focusing the analysis on amusic participants who were unable to discriminate large pitch changes (i.e., pitch discrimination thresholds superior to one semitone). However, Albouy, Caclin et al. (2019) proposed a reanalysis of these data with a multivariate pattern analysis approach. This more sensitive analysis allowed for revealing a difference between amusics and controls in the pattern of functional activity in the right Heschl's gyrus. Linear classifiers based on task-related fMRI data (of this active perception task, but also short-term memory tasks with pitch information, for example) allowed for classifying individuals rather successfully as either amusic or control. These findings contribute to other recent propositions to use task-related imaging data as diagnostic tools for developmental disorders and as predictors of symptom severity.

In Albouy, Mattout, et al. (2013), we measured brain activity during pitch short-term memory tasks, that is tasks directly tapping into amusics' deficits, with MEG recordings. Functional abnormalities were observed during encoding, maintenance, and retrieval phases, all providing converging evidence for anomalies in the involved fronto-temporal networks. During the encoding of the melodies, the MEG measurements revealed decreased and delayed N100m components in bilateral IFG and STG, suggesting not only higher-level processing but also stimulus representations that are abnormal in the amusic brain. Interestingly, N100m alterations can also be observed without a memory task, but just when single isolated tones were presented (i.e., encoding without memory task) (data analysis in progress).

During the short-term memory task with tone sequences, Dynamic Causal Modeling revealed that amusics' alteration were linked with decreased fronto-temporal connectivity, both backward (during encoding) and forward (during retrieval) (see Albouy, Mattout, et al., 2013, 2015). Furthermore, during encoding, the lateral connectivity between the two auditory cortices was increased and the intrinsic modulation within both auditory cortices was decreased. Altered effective connectivity was also observed for retrieval within and between the two auditory cortices.

At this point, it might be argued that functional activity and connectivity differences between amusics and controls may be related to strategies adopted during the experimental tasks. Motivation, attention, voluntary or involuntary compensatory mechanisms of primary deficits due to amusia could explain part of the measured differences. To study this question, resting state data were recorded in a group of amusic participants and their matched controls (Leveque et al., 2016). In the absence of any task and without music, several anomalies were observed. With seeds placed in the primary auditory cortex (Heschl's Gyri), functional connectivity analyses revealed an underconnectivity within the frontotemporal network and an overconnectivity between the two auditory cortices in amusic participants compared with control participants. Furthermore, the auditory cortices were overconnected to the Default Mode Network (DMN). Similar connectivity alterations involving the DMN have been reported in other neurological or psychiatric diseases (Castellanos et al., 2008; Chai et al., 2011), and might be linked with an incomplete maturation of the system (Stevens et al., 2009). Anomalies of the segregation process for the auditory network and/or temporo-frontal networks (see also Hyde et al., 2006) could for instance have interfered with typical neurodevelopment. The resting state data do not allow as such for disentangling causal dysfunctions of congenital amusia from compensatory reorganizations of brain networks secondary to amusia. But they discard strategic behaviors as the only source of anomalies measured in the tests. They suggest that amusic brains are intrinsically different in the way networks are connected, including within the auditory cortex and beyond.

While functional abnormalities have been investigated for music perception and specifically for tone STM, it remained to be shown whether functional abnormalities can also be observed for verbal material (perception, memory). Unimpaired verbal memory performance (e.g., Tillmann et al., 2009; Williamson, Baddeley, et al., 2010) suggests unimpaired functional networks for the processing of verbal material. However, overlapping networks have been reported for both materials (Hickok et al., 2003; Koelsch et al., 2009; Schulze et al., 2011; Schulze & Koelsch, 2012) and could also suggest a higher-level short-term-memory network being affected in amusia not only for tone material, but also verbal material. Albouy et al (2019) conducted an fMRI study where amusics and matched controls performed a delayed matched-to-sample task with tones and words as well as control perceptual tasks. As expected, amusics' performance was impaired for tone perception and memory tasks, but not for the same tasks applied to verbal material (Figure 2), thus replicating the domain-specific short-term memory impairment in amusia (see Tillmann et al., 2009). Functional imaging data during encoding of the tone sequences confirmed decreased connectivity between right STG and right IFG, as previously shown with MEG (Albouy, Mattout, et al., 2013). For the maintenance phase, the comparison between tonal and verbal material was particularly interesting as no signal was presented during the delay (here nine seconds). For the verbal material, the participant groups did not differ, both showing activation of the left IFG as well as increased fronto-temporal connectivity between the left IFG and the left anterior STG in the memory task (in comparison to the control perception task). However, for the tone material, amusic participants showed decreased activation of right frontal (IFG, DLPFC), right temporal regions, and left IFG as well as increased activation of some auditory regions. These results suggest that during tone maintenance, amusics recruited brain areas encompassing mainly auditory regions, which reveals an inefficient strategy, notably with left STG activity correlating negatively with memory performance. This contrasts with controls' results, with right IFG activation correlating positively with memory performance.

In addition to these alterations confirming abnormal fronto-temporal pathways, the connectivity between right IFG and right DLPFC was decreased in amusics during maintenance (in comparison to controls) for the tonal material in this fMRI study (Albouy et al., 2019). This finding is in agreement with results of a gamma-band activity analysis in the MEG study (Albouy, Mattout, et al., 2013), showing that while controls recruited the right DLPFC during the maintenance delay of the tone short-term memory task, amusics did not. The role of the right DLPFC in tonal maintenance is in line with Schaal et al. (2015), showing that the modulation of this region with 35Hz (gamma) transcranial Alternating Current Stimulation causally improves pitch memory performance in amusic participants. These observations related to the deficit have received converging evidence by the study of Royal et al. (2018) using tDCS with stimulation over frontal regions in typical, non-amusic participants, thus creating artificially an amusic profile. More recently, Samiee et al. (2022) reported congruent deficit patterns in amusia for a pitch change detection task, by analyzing the role of oscillatory brain activity and, in particular, between brain regions with specific cross-frequency dynamics. For amusia, the findings revealed altered phase-amplitude coupling in the auditory cortex together with decreased inter-regional signals to inferior frontal cortices and to motor cortices. They provide further insights into amusics' anomalies, notably suggesting a misalignment between stimulus encoding and predictive timing of the auditory material.

The overall data pattern is thus that of an impaired fronto-temporal network in congenital amusia, particularly in the right hemisphere, with furthermore anomalies in the interactions between auditory networks and other large cortical networks (DMN, motor networks). Importantly, the deficit starts as early as the auditory cortex and might also affect auditory processing in the brainstem (Lehmann et al., 2015; but see Liu et al., 2014, for conflicting evidence).

5. From fronto-temporal pathway anomaly to the hypothesis of potential disorder of consciousness

In addition to the research showing the anomalies of the fronto-temporal network, involved in pitch encoding, maintenance, and retrieval, has emerged the hypothesis of a potential disorder of consciousness in amusia, that is abnormalities in the conscious access of tonal information processing (e.g., Omigie & Stewart, 2011; Peretz et al., 2009). This hypothesis is motivated by previous findings for various neurological disorders (e.g., aphasia, prosopagnosia). Indeed, numerous research has shown the power of implicit perception and cognition. Implicit processing can remain functional despite the disorder, contrasting with impaired or decreased processing capacities on an explicit, conscious level. Indirect investigation methods have revealed spared implicit processing capacities in the presence of severe impairments in tasks requiring explicit processing. This has been observed not only in classical cases affecting visual perception and language processing (e.g., Mimura et al., 1996; Schacter & Buckner, 1998), but also in a case of acquired amusia (Tillmann et al., 2007).

Both behavioral and EEG findings have provided evidence that individuals with congenital amusia might have less impaired pitch and tonal structure processing than previously suggested with tasks requiring explicit judgments. The present section provides an overview of these findings using indirect investigation methods and revealing some spared pitch processing capacities in congenital amusia as well as some tonal knowledge, notably about the tonal structures of the musical system (i.e., tonal enculturation), and about specific culturally familiar repertoires stored in long-term memory (also referred to as « musical lexicon », Peretz & Coltheart, 2003). Further investigations of potentially remaining, intact functions in congenital amusia have also implications for boosting musical short-term memory (see section 6, Albouy, Schulze, et al., 2013a; Lévêque et al., 2022) and for perspectives for training and rehabilitation, notably by aiming for training that can build on or exploit spared implicit processing resources (e.g., Kessels & deHaan, 2003).

At the neurophysiological level, some spared pitch processing was shown thanks to EEG measurements that revealed an automatic brain response following pitch anomalies in musical sequences, anomalies that congenital amusic participants were not able to report explicitly (Peretz et al., 2009). However, such a preserved response was observed only at an early latency (around 200 ms after the pitch anomaly), whereas later responses observed in controls (around 600 ms) were absent in amusics (in keeping with Zendel et al., 2015). This early automatic response was observed for out-of-tune tones and not out-of-key tones, revealing some insensibility to the larger musical context. This finding of preserved early automatic brain responses to pitch anomalies in musical sequences is in agreement with the fact that the MMN after pitch changes within sequences of repetitive standard sounds is preserved to some extent in congenital amusia (Moreau et al., 2013; Quiroga-Martinez et al., 2021; Quiroga-Martinez et al., 2022), with however a markedly reduced P3a after the MMN (Moreau et al., 2013; Pralus et al., 2020). Focusing not on pitch changes, but on how predictable each note within a musical phrase was, Omigie et al. (2013) reported that congenital amusics retained a sensibility to the expectedness of each note, yet not as pronounced as that of controls. These EEG results suggest some spared implicit capacities of the amusic brain to process the pitch dimension, at least with less impairment than suggested by behavioral result patterns based on explicit judgments of the musical material (see section 2). They have further motivated the hypothesis that thanks to these implicit pitch processing capacities, amusic participants might still be able to acquire some musical, tonal structure knowledge about the musical system of their culture, even though potentially more sparse than what has been shown for the non-amusic, nonmusician population (e.g., Bigand & Poulin-Charronnat, 2006).

One indirect behavioral investigation method that has been used with congenital amusics is the priming paradigm (Omigie et al., 2013; Tillmann et al., 2012). This paradigm, extensively used in psycholinguistics (e.g., Neely, 1991), was introduced into the music cognition domain by Bharucha and Stoeckig (1986) for chord pairs and has been further developed for chord sequences and melodies to investigate nonmusicians' tonal knowledge (e.g., Bigand & Pineau, 1997; Marmel et al., 2010). The central feature of the priming paradigm is the indirect investigation of the context's influence on listeners' expectations and event processing. Participants are not required to make direct judgements on the relation between prime context and target, but their task focuses on a perceptual feature of the target chord (or tone), such as its consonance/dissonance, the used timbre or sung syllable. The influence of tonal structures and expectations is shown by processing speed differences of the musical target events (i.e., faster processing of musically related targets, which should be more strongly expected, than of unrelated or less-related (unexpected or less-expected) ones). Omigie et al. (2012) adapted this paradigm to melodies and showed that the probability of occurrence of tones in melodies (i.e., linked to listeners' expectations) influenced tone processing similarly for both amusic and control participants. However, when the same material was used with explicit judgements of the expectedness of the target notes, amusics were impaired in comparison to controls. While Omigie et al. (2012)'s material included variations not only in terms of tonal structures, but also on other features, such as the melodic contour, Tillmann et al. (2012) provided converging evidence for amusics' implicit tonal processing by focusing on tonal-harmonic structures only. As previously observed for nonmusicians (adults, children), amusic and control participants processed faster the structurally more important (supposed to be more strongly predicted) target chords than the less important ones. Even though the difference was less pronounced for the amusic participants than their control participants, this finding suggests that amusics have acquired at least some structural knowledge about the musical system of their culture.

Amusics' implicit knowledge about the structures respecting the Western tonal musical system stored in long-term memory has been further shown in a study investigating amusics with three questions judging tonal and atonal versions of musical pieces. Amusics' judgments showed that they could discriminate between the two versions with all three questions, but the extent of the revealed tonal structure processing was influenced by the task demands. While amusics were impaired for a question requiring explicit structural judgments of the musical pieces (i.e., in reference to the musical system of their culture), they performed as well as their matched controls for two other questions that focused on either a more personal, emotional dimension (i.e., buying a CD with this music) and a more social one (i.e., judging the perception of others, notably for ranking of the pieces in a French hitparade). Interestingly, the influence of task demands has also been shown for the EEG measurements reported in Zendel et al. (2015). Amusics' evoked potentials reflected their detection of out-of-key tones only when the task required a judgment on a different dimension i.e., the detection of unrelated clicks), but not when the task required the explicit detection of the out-of-key tones.

The implicit processing capacities of amusics have led not only to tonal structure knowledge stored in long-term memory, but also to the storing of frequently encountered ("familiar") music in long-term memory (musical lexicon). While amusics have reported their difficulty in recognizing melodies without lyrics, first indirect evidence for some musical lexicon can be found in Ayotte et al. (2002). For both amusics and controls, performance in a pitch anomaly detection task was better for familiar melodies than for unfamiliar melodies. Converging evidence was reported by (Quiroga-Martinez et al., 2021) showing larger MMN to pitch changes in familiar than in unfamiliar melodies for amusic participants. Further evidence for long-term musical memories in amusia was obtained in a recent study investigating the phenomenon of musical "earworms" (Tillmann et al., 2023). Almost all amusic participants reported to experience musical "earworms", which relate to involuntary memory reactivations. Yet amusic participants reported to do so less often than control participants, thus suggesting that long-term memories or access to them might differ between groups.

Using an explicit familiar melody recognition task based on a closed set of possible melodies, Graves et al. (2019) showed that amusic participants performed significantly above chance level, even though they were impaired relative to controls. When participants were required to only judge the degree of the evoked feeling of familiarity without explicit recognition, amusics' response patterns did not differ from the ones of controls. Using a gating paradigm with an open-set testing, Tillmann et al. (2014) investigated the minimal amount of acoustic information necessary to access long-term knowledge about familiar music. Participants provided familiarity judgments for segments of familiar and unfamiliar instrumental musical pieces, which were presented with increasing duration (starting with excerpts of 250ms, then 500 ms, 1000ms etc). Results revealed that amusics were able to perform the task consistently over time (i.e., with increasing duration) and that their familiarity judgments differentiated familiar from unfamiliar excerpts starting with 500ms-duration excerpts, as observed for the controls. The response pattern between the groups differed only for the response times in reaching the judgments. For the longer excerpts, amuiscs responded overall more slowly than did the controls, and for the shorter excerpts amusics' response times suggested to need more time to reach their judgments for familiar excerpts. These findings thus showed that amusics have built up a musical lexicon, even though they might have a slower access or need additional processing time because of their uncertainty and/or low confidence in their music perception abilities.

Investigating amusics' potential production abilities might provide further understanding of the extent of impairments, but also of preserved processes in amusia, including the contribution of auditory-motor feedback and of a vocal motor code (Hutchins & Peretz, 2013). In particular, amusics can imitate what they cannot discriminate in perception, also suggesting implicit competences of pitch processing in amusia (e.g., Hutchins & Peretz, 2012; Loui et al., 2008).

Taken together, these findings further support the hypothesis that congenital amusia might be related to some impaired conscious access to music processing rather than music processing per se, even though more research is needed to further investigate this phenomenon and its underlying neural correlates, which is possibly related to the fronto-temporal loop. Measuring resting state activity, Jin et al. (2021) reported abnormalities also in the posterior cingulate and the precuneus, which can also be linked to consciousness.

6. <u>Boosting pitch encoding and pitch memory in amusia</u>

Congenital amusia has previously been described as a persistent, life-long condition, but recent research suggests that lasting improvement might be achieved with training. Some first attempts at rehabilitation showed encouraging results after 7 weeks of singing workshops (Anderson et al., 2012), but no success after 4 weeks of daily music listening in children (Mignault Goulet et al., 2012). More recently, a study involving psychophysical training that targeted fine-grained auditory processing demonstrated improvement for amusics (Whiteford & Oxenham, 2018). Amusics and controls spent four sessions (i.e., 1-2 hours each) performing a basic psychophysical task, either pitch discrimination or interaural level discrimination, and completed the MBEA before and after training. Both groups improved in their MBEA scores, regardless of the trained stimulus, and over half of the trained amusics no longer qualified as amusic after training, even maintaining the same level of performance one year after training. It is relevant to note, however, that the authors acknowledge to be unable to directly identify whether the improvements on the MBEA observed in amusics were associated to i) a test-retest effect, ii) to the learning caused from the 4-session psychophysical training (generalization to untrained stimuli), or iii) a combination of the two. Nevertheless, the authors elegantly toned down the potential contribution of test-retest effects by referring to the results of Liu et al. (2017) in which no test-retest effects have been observed in a separate group of untrained amusics (in the pitch-subtask of the MBEA - with a test-retest gap of 2 weeks, comparable to the temporal gap used in Whiteford & Oxenham, 2018). This study thus supports further the interpretation that the psychophysical training performed in Whiteford and Oxenham (2018) might have generalized to improve melody and/or pitch discrimination abilities in the trained amusics. Together with Anderson et al. (2012), this study represents encouraging perspectives for rehabilitation of amusia.

In the following, we report other research that aimed to identify what could facilitate pitch encoding and support pitch memory in congenital amusia. More specifically, the studies target (1) characteristics of the material (such as duration, speed of presentation or complexity), (2) its simultaneous presentation with visual stimuli to benefit from cross-modal integration as well as benefits based on (3) listeners' long-term memory knowledge, notably either on the musical structure (i.e., tonality) or on specific musical pieces (i.e., musical lexicon), and on (4) liking.

Material characteristics. One factor that has been shown to reduce the deficit for short-term pitch memory in amusia is increased duration of tones (Albouy et al., 2016). This deficit might be explained not only in terms of reduced backward masking, but also to increased reliance on temporal envelope cues in amusia. In listeners without amusia, pitch discrimination thresholds for pure tones improve with increased tone duration (Moore, 1973), up to 200 ms and especially for frequencies below 5 kHz, likely due to the limit of phase locking. This suggests that increased temporal envelope information with longer tone durations is helpful, especially in the spectral region where this cue is available. In general, amusics seem to exhibit greater reliance on non-spectral cues, such as amplitude modulation, loudness, and temporal-envelope-based pitch of unresolved harmonics (Cousineau et al., 2012, 2015; Graves et al., 2019b, 2021). Indeed, when temporal envelope and spectral information are in opposition to each other, with "chimera" stimuli, amusics exhibit a greater reliance on temporal envelope cues (Bones & Wong, 2017). Future studies need to further investigate pitch perception in amusia using stimuli that provide varying levels of access to temporal envelope cues, and focused rehabilitation programs might make use of stimuli where these cues are exaggerated or easy to use, such as tones with longer duration.

  <u>Audio-visual stimulations.</u> To perceive environmental stimuli, multisensory interactions are essential. For instance, the McGurk effect shows that the integration of visual and auditory information assists speech perception (Mcgurk & Macdonald, 1976). At the brain level, several pieces of evidence have been reported that interactions are present across sensory modalities (Shams & Seitz, 2008; Shimojo & Shams, 2001). Multisensory integration has been reported to be stronger when one of the sensory modalities is deficient (Frassinetti et al., 2005; Grasso et al., 2016; Passamonti et al., 2009). For example, in participants with reduced visual acuity, auditory cues presented simultaneously with visual cues (and resulting audiovisual interactions) allowed improving their visual detection threshold (Gabor patches) beyond their visual only performance, which was not observed in control participants (Caclin et al., 2011). Similarly in participants with congenital amusia, visual stimulations improved performance in an auditory pitch change detection task (Albouy, Lévêque, et al., 2015). In this task, the visual information was uninformative regarding the pitch detection task, but provided temporal cues about when the onset of the tone occurred. Amusics demonstrated audiovisual integration abilities similar to control participants and their response times were significantly shorter even for small pitch intervals, which were not correctly detected without visual cues.

  These benefits of cross-modal integration lay out some possibilities of remediation of pitch deficits in congenital amusia (see also Lu et al., 2016, 2017). To investigate more specifically the potential benefit of multisensory integration in congenital amusia, a rehabilitation program using audiovisual tasks was designed in our lab (in progress). This training is composed of three pitch specific tasks, a pitch change detection task, a pitch direction change identification task and a short term memory task, with half of the auditory trials being presented with either informative or non-informative visual stimuli. Nineteen controls and eighteen amusics performed this training over fifteen weeks, with two sessions of 20 minutes per week. We used a visuospatial training as a control training (Bedoin & Médina, 2013). All participants underwent both the audiovisual and the visuospatial training in a cross-over design, with the order of the trainings counterbalanced across participants. In order to investigate the effect of training on brain plasticity, MEG measures were recorded during a pitch short-term memory task and a passive oddball paradigm before and after the training sessions. Preliminary results revealed that amusics seems to benefit from audiovisual pitch training to identify pitch direction changes, leading to higher accuracy and shorter response times after the audio-visual training, but not the control training (thus excluding an explanation in terms of test/retest benefit). The MEG data revealed a larger MMN to small pitch changes (0.25 semitone) in amusics after the audiovisual training. These results suggest that amusics could benefit from multisensory integration to improve pitch processing. Further analyses are currently in progress to evaluate the effect of this pitch audiovisual training on pitch encoding and whether it could induce changes in neural correlates (e.g., fronto-temporal network).

Long-term memory knowledge. Beyond perceptual training in either the auditory modality (Whiteford & Oxenham, 2018) or exploiting audio-visual interactions as well as singing training (Anderson et al., 2012; see also Wilbiks et al., 2016), amusics' pitch memory performance can also benefit from knowledge stored in long-term memory, as previously shown for non-amusic, non-musician listeners. This benefit can be based on listeners' implicit knowledge of the structure of the musical system (i.e., tonality) or on listeners' knowledge of specific musical pieces stored in the musical lexicon (i.e., familiar music). Amusic's long-term musical knowledge has been supported by reports of musical earworms (i.e., involuntary musical imagery) in congenital amusia (Tillmann et al., 2023). These findings further suggest the possibility to also build on long-term memory and mental imagery as other training or rehabilitation strategies.

Tonality provides structural cues that can improve short-term memory. Even nonmusician listeners show better memory performance for tonal melodies or tonal chord sequences than for atonal version thereof, which are devoid of tonal structure (e.g., J.

Bharucha, 1983; Dowling, 1991; Schulze et al., 2012). Using 5-tone sequences, we replicated this benefit of tonal structure for short-term memory (as measured by d') for the control participants and extended this to response times. While amusics' memory performance did not differ between tonal and atonal material, amusics showed a benefit of the tonal structures in their response time patterns (Albouy, Schulze, et al., 2013a). In a second study, we implemented the tonal versus atonal versions in orchestrated, longer musical excerpts, thus richer material with more tonality cues (Lévêque et al., 2022). With this material, amusics' STM performance (in terms of d') was still impaired, but it was improved for tonal over atonal materials, as observed for control participants. This beneficial effect of tonality on short-term memory reflects implicit knowledge stored in long-term memory in amusia. Even amusic individuals seem to be able to build a representation of these musical structures and regularities (see Section 5), and this long-term knowledge is mobilizable to support deficient memory processes.

As summarized in section 5, amusics also have stored specific musical pieces in long-term memory. Despite their subjective reports about explicit recognition difficulties, they can report feelings of familiarity (Tillmann et al., 2014) and can recognize familiar melodies above chance in a closed-set paradigm (Graves et al., 2019). Retrieval of melodies stored in LTM was possible for amusics when the melody was presented using pitch and when using other acoustic dimensions (Graves et al., 2019), suggesting that the main LTM difficulty may be in retrieval and not in encoding. The possibility that underlying LTM storage of melodic contours may remain intact in amusia and facilitates even pitch processing (see Section 5, e.g., Ayotte et al., 2022; Quiroga-Martinez et al., 2021) is encouraging for potential rehabilitation and might be worth exploring to help training STM for pitch.

Liking. Previous research investigating musical memory in non-amusic (either nonmusician or musician) populations has shown that listeners remember better music they like (e.g., Stalinski & Schellenberg, 2013; Ferreri et al., 2021). We investigated this potential benefit of liking on musical long-term memory in congenital amusia, and replicated the previous result even in this population. Musical excerpts rated as "liked" (rated 4 or 5 on a liking scale from 1 to 5) were significantly better recognized in the second part of the testing session when presented among foils (Lévêque et al., in revision). The musical stimuli were unfamiliar, but taken from real-world musical recordings. This demonstrated an influence of personal musical appreciation on memory, even when this memory is impaired as it is in amusia. The memory network thus appears to benefit from connections with emotional networks. Rehabilitation could thus take advantage of emotions to improve impaired cognitive processes as music memory.

This set of results suggests that, despite the vulnerability of the memory trace for music in amusia, which is highly sensitive to memory load, interference, or time, the auditory memory network benefits from connections with other networks like visual, emotional, or long-term memory networks. This sketches perspectives for training and rehabilitation, taking advantage of these preserved resources to improve impaired perceptual and cognitive processes.

7. Conclusion and Perspectives

This review presents empirical data on the rare condition of congenital amusia. Albeit known since a long time (Allen, 1878), it has been scientifically studied solely since about twenty years (starting with Peretz et al., 2002). The research provides better understanding of the condition itself, but also of normal perceptual and cognitive functioning and its underlying neural correlates. Congenital amusia has been previously labeled as "musical handicap" (Peretz & Hyde, 2003) and presented as a unique opportunity to study the interplay between behavior, brain, genetics and environment (e.g., Peretz, 2016). It has been suggested to be a life-long-deficit and referred to as neurodevelopmental disorder, leading to comparisons to other neurodevelopmental disorders, such as dyslexia or prosopagnosia (Corrow et al., 2019; Couvignou et al., 2019; Couvignou & Kolinsky, 2021). The hypothesis of being a life-long deficit with genetic bases now calls for further research, in particular to reinforce research investigating this condition in childhood (e.g., Couvignou & Kolinsky, 2021; Lebrun et al., 2012; Peretz et al., 2013; Wilcox & Biondi, 2015), including for rehabilitation perspectives, and to combine its investigation with current attempts to investigate the genetic bases of musicality (e.g., Honing, 2018). Further investigations of potential comorbidity with other neurodevelopmental disorders, such as dyslexia (e.g. Couvignou et al. 2019, 2023; Couvignou & Kolinsky, 2021), provides new perspectives for the understanding of neurological and neurodevelopmental disorders, such as common impaired mechanisms (e.g., perceptual and cognitive sequencing) as well as the potential use of musical material for training and rehabilitation.

Current findings have focused mainly on the pitch deficit in amusia, whether for perception or memory. They have provided further insights in STM functioning, implicit processing, and also the potential domain-specificity or generality, notably by comparing tonal versus verbal material. The altered neural correlates in amusia have provided further insights in fronto-temporal network relevance for pitch encoding and memory as well as the connectivity with homologous temporal regions in the other hemisphere and intrinsic modulation in the auditory cortex. Starting from structural and functional description with localized regions, co-occurrence and co-activation of regions, more recent analyses have investigated both functional and effective connectivity networks (e.g., Albouy, Mattout, et al., 2015; Albouy, Peretz, et al., 2019; Hyde et al., 2009) and now getting to the investigation of more complex brain network dynamics, such as oscillatory activity, cross-coupling across frequencies and regions (see Samiee et al., 2022 for a first step). Samiee et al. (2022) showed that the involved impairments in auditory and frontal networks extend to motor regions and connections. These findings thus allow for combining the involved networks to those observed in predictive timing and predictive coding frameworks (e.g., Arnal & Giraud, 2012) including those related to active sensing (e.g., Morillon et al., 2019). They now open up to further suggest the hypothesis that amusia might be related to a more general perceptual and cognitive sequencing deficit, which expresses in particular for material that does not allow for explicit verbalization strategies. In addition, future research also needs to study the role of frontal regions (inferior frontal and DLPFC) in pitch short-term memory, including the potentially altered access to consciousness, such as via altered frontal activity or connectivity.

A final promising research direction that the domain of congenital amusia is currently taking is the investigation of potential perspectives for training and rehabilitation. These attempts now could further build on the acquired understanding of spared processes, the benefit of inter-modal integration, some enhanced perceptual features, long-term knowledge as well as implicit processes and liking. For these directions, as well as for the investigation of involved neural correlates, future research should further consider that congenital amusia might be a phenomenon with multiple profiles, expressing itself differently and altering pitch, time, or emotional dimensions of music processing with different weighting. Adapting the labeling of "congenital amusias" might be a way to explicitly remind this potential multiple-profile phenomenon (see also Tillmann et al., 2015). Investigation of amusia could benefit also from a more systematic assessment of perceptual, cognitive, and emotional aspects, as has been proposed by Stewart et al. (2006) for acquired cases and with extension to congenital ones.

**Figure Captions**

Figure 1.

MBEA scores and Pitch Discrimination Threshold (PDT) data for amusic (n=69) and control (n=175) participants tested in Lyon. Groups were created as follows to avoid borderline cases: amusic participants had a global MBEA score below 23 (average across the six subtests of the MBEA, maximum score = 30), and control participants had a global MBEA score above 24.5. We report for each group (amusics' data in red, controls' data in blue) the global MBEA score (Panel A), the MBEA pitch score (average of the performance at the first three subtests, Panel B), the MBEA rhythm score (Panel C), PDT (procedure from Tillmann et al., 2009; Panel D). Panels E and F present the MBEA pitch scores for each group (amusics in panel E and controls in Panel F) as a function of the participants PDT, below or above one semi-tone. In each panel, each point represents data from a single participant, the whisker plots illustrate the median and interquartile range, and finally a smoothed histogram illustrates the distribution of the data.

Figure 2.

Performance of amusic (n=18) and control (n=18) groups  in terms of d', presented as a function of material (tonal; verbal) and task (memory task; perception task). Data from Albouy et al. (2019), presented as the MBEA and PDT data in Figure 1. The memory task was a delayed-matching-to-sample task, with sequences of 3 items to memorize, with a silent retention delay of 9 seconds. The perception task simply consisted in comparing two consecutive items in a same-different judgment task.

Figure 3.

A. Overall patterns of cortical anomalies in congenital amusia, synthesizing results from anatomical and functional studies, in particular VBM data (Hyde et al., 2006), DTI studies (Chen et al., 2018; Loui et al., 2009; Wang et al., 2017), MEG data (Albouy, Mattout, et al., 2013, 2015), fMRI data (Hyde et al., 2011; Lévêque et al., 2016; Albouy et al., 2019). See main text for details. AC: Auditory Cortex; IFG: Inferior Frontal Gyrus; DLPFC: Dorso-Lateral Prefrontal Cortex; DMN: Default Mode Network.

B. Specific anomalies in the right fronto-temporal network during short-term memory. The

main abnormalities observed during each stage of memory processing (encoding, maintenance, and retrieval) are highlighted (data from Albouy, Mattout, et al., 2013, 2015; Albouy et al., 2019).

**Keypoints:** Overview of main features regarding congenital amusia as summarized here

**Insert 1**

**Core deficits in congenital amusia**
Deficits in pitch processing including:
- elevated pitch discrimination thresholds (including for gliding pitch sounds)
- impaired pitch direction and contour processing
- impaired short-term memory of pitch sequences (melodies) that can be observed even without elevated pitch discrimination thresholds

**Insert 2**

**Different subtypes of amusia - Pitch processing deficits can be accompanied or nor by:**
- Rhythm and beat processing deficits
- Reduced music emotion and enjoyment
- Reduced music seeking and daily use
- Poor singing in most cases but not all

**Insert 3**
**Factors boosting music perception and memory in congenital amusia:**
i. Tone material features and presentation
   Reducing speed of presentation: increasing tone duration or inter-tone intervals
   Psychophysical training targeting fine-grained auditory processing
ii. Audio-visual presentation
   Visual cue indicating the onset of tones (temporal cue)
   Audio-visual training (pitch change detection, pitch direction, pitch short term memory)
iii. Listeners' long-term memory knowledge
   Presence of musical structure (i.e., tonality)
   Reference to listeners' musical lexicon (familiarity)
iv. Liking
   Liked music is better remembered

**Figure 1.**

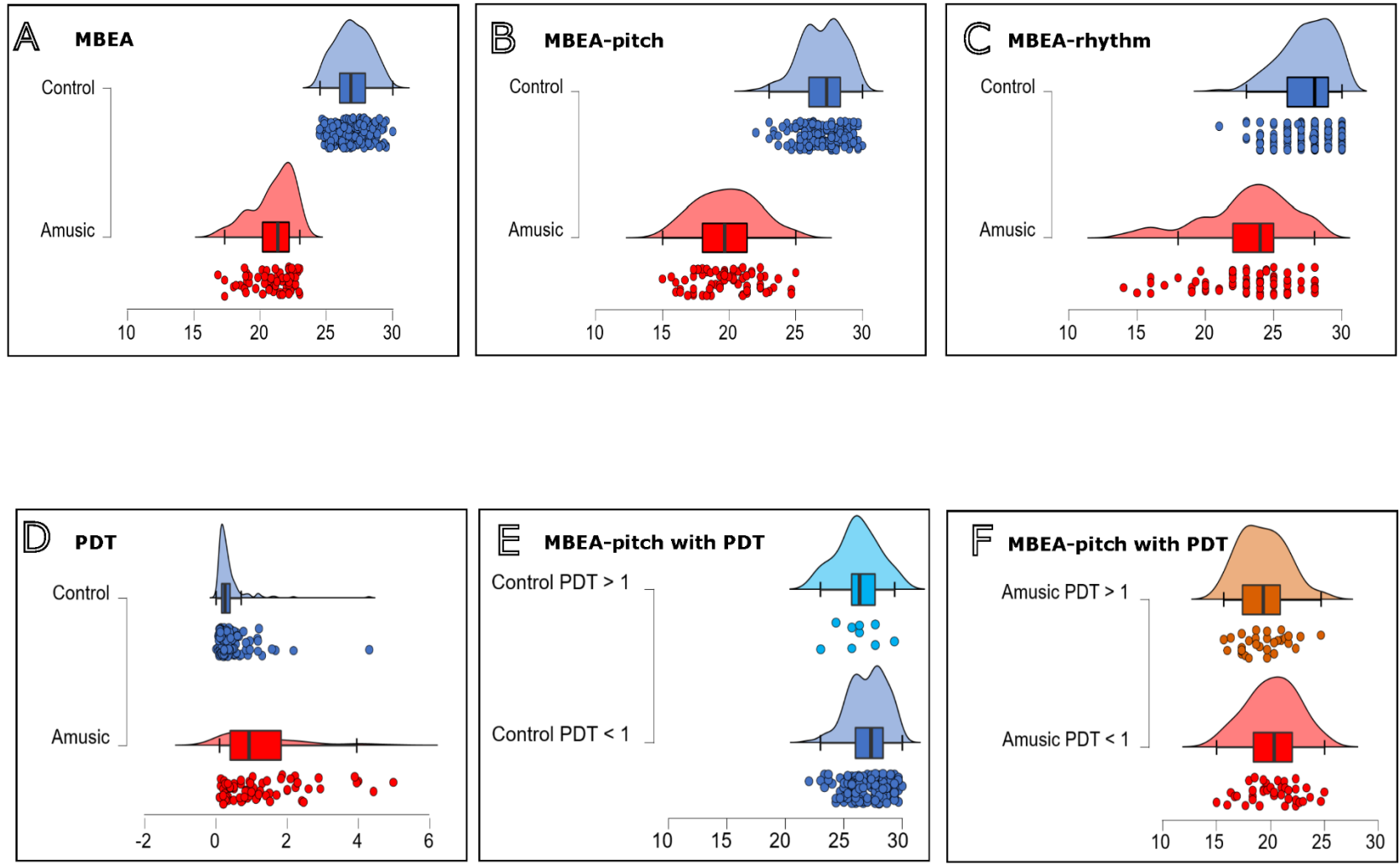

**Figure 2.**

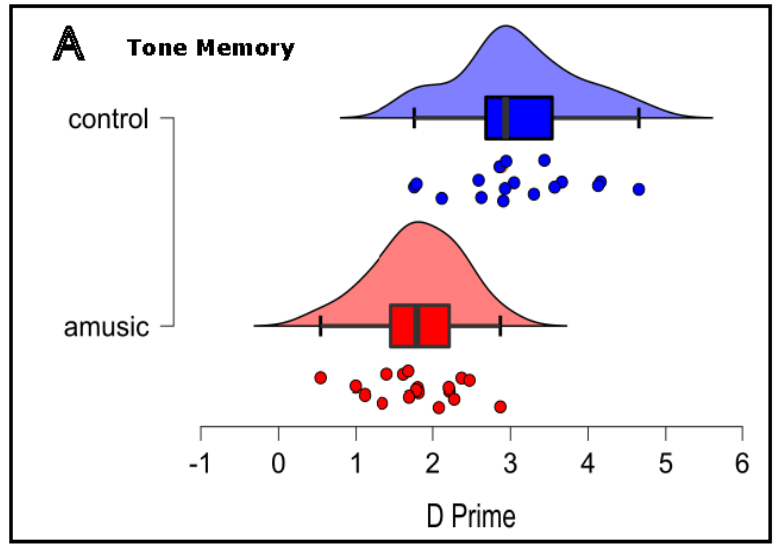

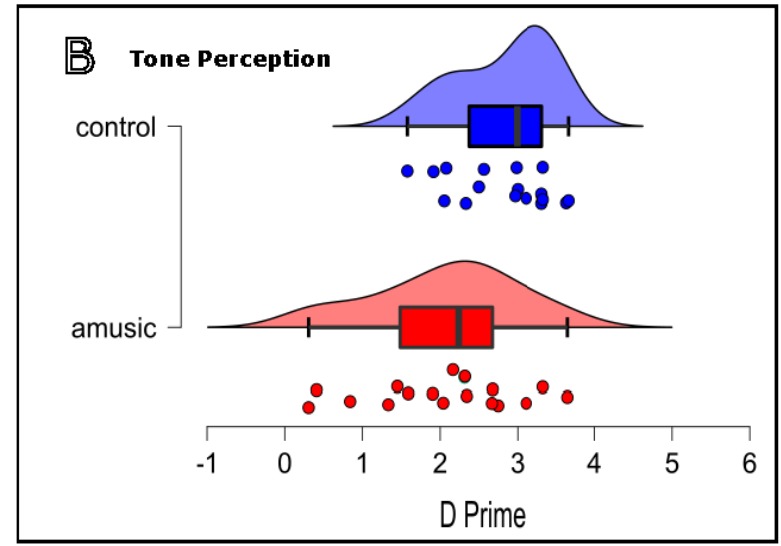

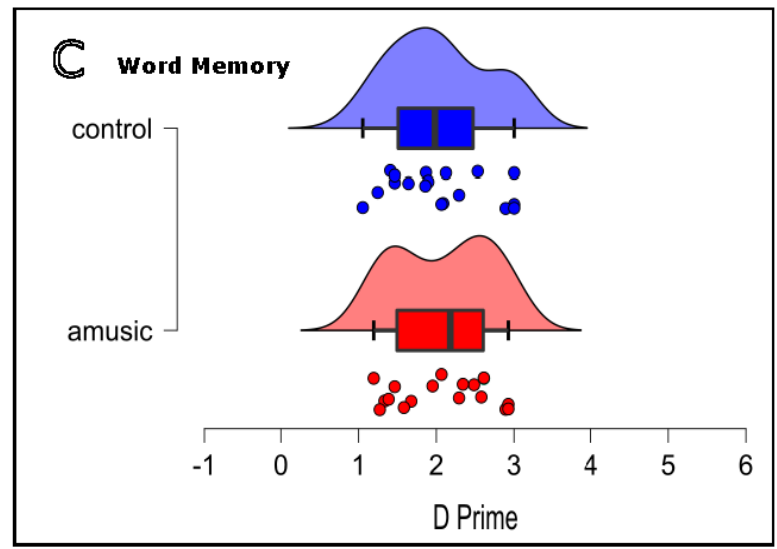

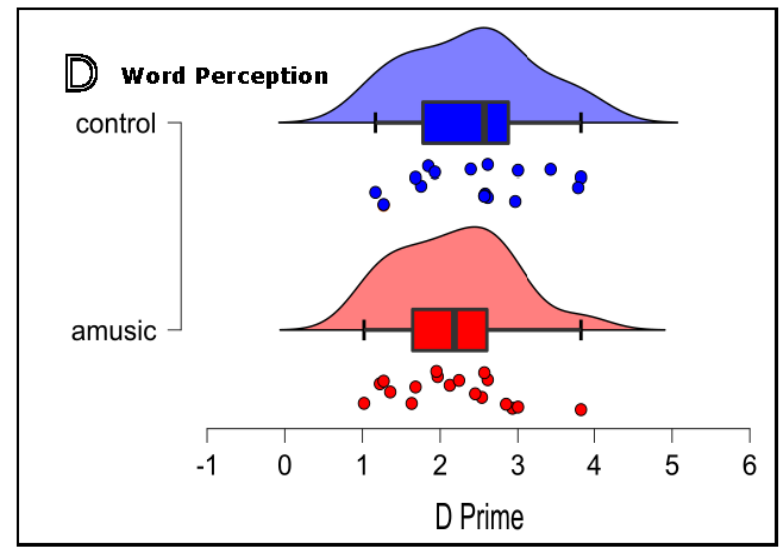

**Figure 3.**

A. Overall pattern of cortical anomalies in congenital amusia

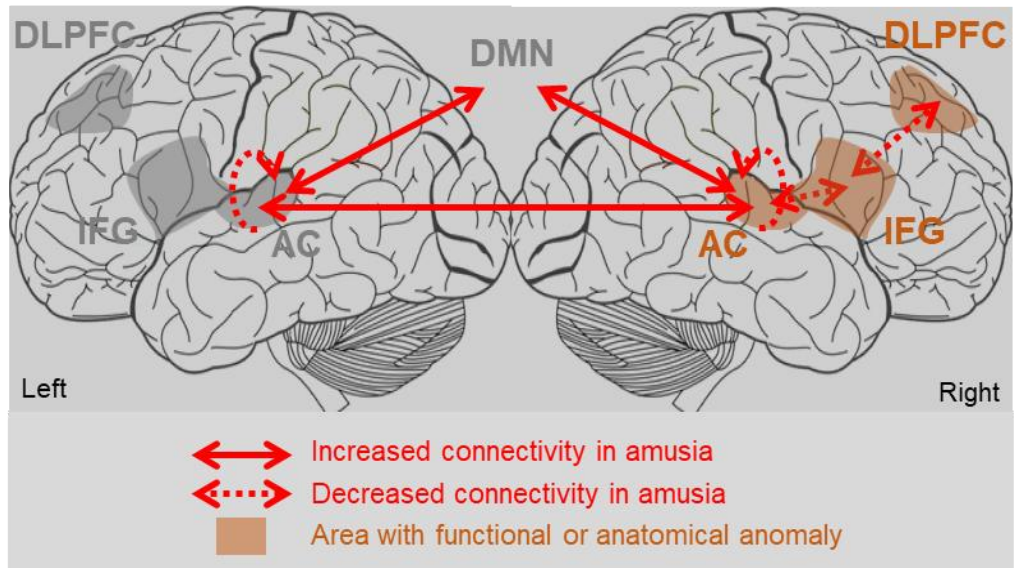

Left

Right

➡️ Increased connectivity in amusia

◀┈┈▶ Decreased connectivity in amusia

⬛ Area with functional or anatomical anomaly

B. Focus on pitch short-term memory

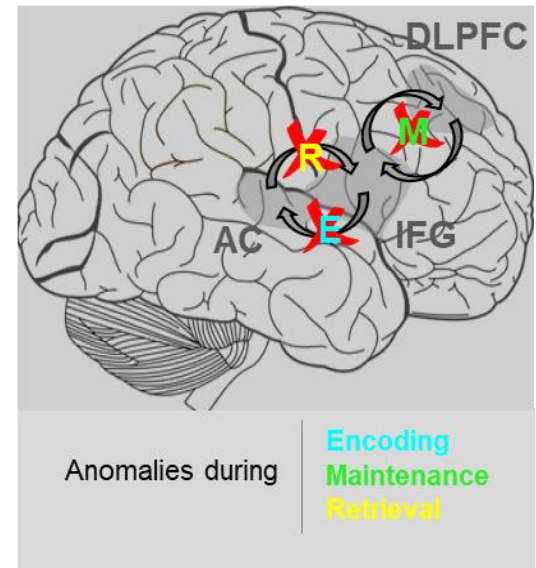

Anomalies during

Encoding
Maintenance
Retrieval